\documentclass[runningheads,a4paper,floatfix]{llncs}

\usepackage{subcaption}
\usepackage{amsmath}
\usepackage{amssymb}
\usepackage{listings}
\usepackage{multirow}
\usepackage{fancyvrb}
\usepackage{xcolor}

\usepackage{xspace}
\usepackage{tikz}
\usepackage{cite}
\usepackage{pgfplots}

 \usepackage{hyperref}

\newcommand{\figref}[1]{Fig.~\ref{Fi:#1}}
\newcommand{\figrefs}[2]{Figs.~\ref{Fi:#1} and~\ref{Fi:#2}}
\newcommand{\defref}[1]{Defn.~\ref{De:#1}}

\renewcommand{\eqref}[1]{Eqn.~(\ref{Eq:#1})}

\newcommand{\sectref}[1]{Section~\ref{Se:#1}}
\newcommand{\sectrefs}[2]{Sections~\ref{Se:#1} and~\ref{Se:#2}}

\newcommand{\algref}[1]{Alg.~\ref{Alg:#1}}
\newcommand{\algline}[1]{(Line~\ref{#1})}
\newcommand{\alglines}[2]{(Lines~\ref{#1}--\ref{#2})}

\newcommand{\inangleb}[1]{ \ensuremath{\langle #1 \rangle}}
\newcommand{\insquareb}[1] {\ensuremath{\left[ #1 \right]}}
\newcommand{\inroundb}[1]{\ensuremath{ \left( #1 \right)}}
\newcommand{\inparan}[1] {\ensuremath{ \{ #1 \}} }
\newcommand{\vbars}[1]{\ensuremath{\left\vert #1 \right\vert}}
\newcommand{\SC}[1]{\textsc{#1}}

\newcommand{\IT}[1]{\textit{#1}}
\newcommand{\TT}[1]{\texttt{#1}}
\newcommand{\MI}[1]{\ensuremath{\mathit{#1}}}
\newcommand{\MT}[1]{\ensuremath{\mathtt{#1}}}

\newcommand{\MC}[1]{\ensuremath{\mathcal{#1}}}
\newcommand{\BF}[1]{\textbf{#1}}
\newcommand{\EM}[1]{\emph{#1}}
\newcommand{\Omit}[1]{}

\newcommand {\PO}{\ensuremath{<_{po}}}

\newcommand{\cbmc}{\textsc{Cbmc}\xspace}
\newcommand{\dfence}{\textsc{Dfence}\xspace}
\newcommand{\memorax}{\textsc{Memorax}\xspace}
\newcommand{\remmex}{\textsc{Remmex}\xspace}
\newcommand{\minisat}{\textsc{Minisat 2.2.0}\xspace}

\newcommand{\FI}{\textsc{Fi}\xspace}
\newcommand{\robmc}{ROBMC\xspace}
\newcommand{\TE}{\textsc{Te}\xspace}
\newcommand{\robmcet}{\textsc{ROBMC-Et}\xspace}
\newcommand{\tuple}[1]{\langle #1 \rangle}

\definecolor{highlighter}{rgb}{0.85,0.85,0.85}
\newcommand{\highlight}[1]{\colorbox{highlighter}{#1}}

\makeatletter
\newbox\sf@box
{\def\sf@one{#1}%
\def\sf@two{#2}%
\setbox\sf@box\hbox
\bgroup}%
{ \egroup
\ifx\@empty\sf@two\@empty\relax
\def\sf@two{\@empty}
\fi
\ifx\@empty\sf@one\@empty\relax
\subfloat[\sf@two]{\box\sf@box}%
\else
\subfloat[\sf@one][\sf@two]{\box\sf@box}%
\fi}
\makeatother

\usepackage{algorithm}
\usepackage{algorithmic}

\newtheorem{mydefinition}{Definition}

\newif\ifhenabled

\urldef{\mailsa}\path|{saurabh.joshi,daniel.kroening}@cs.ox.ac.uk|

\begin{document}

\mainmatter

\title{Property-Driven Fence Insertion using \\
       Reorder Bounded Model Checking\thanks{This
research was supported by ERC project 280053 and
by the Semiconductor Research Corporation (SRC)
project 2269.002.}$^\ddagger$}


\titlerunning{Property-Driven Fence Insertion using ROBMC}

\author{Saurabh Joshi \and Daniel Kroening}


\institute{Department of Computer Science \\
	University of Oxford, UK \\
\mailsa }

\toctitle{}
\tocauthor{}
\maketitle
\begin{abstract}
Modern architectures provide weaker memory consistency guarantees than
sequential consistency.  These weaker guarantees allow programs to exhibit
behaviours where the program statements appear to have executed out of
program order.  Fortunately, modern architectures provide memory barriers
(fences) to enforce the program order between a pair of statements if
needed.  Due to the intricate semantics of weak memory models, the placement
of fences is challenging even for experienced programmers.  Too few fences
lead to bugs whereas overuse of fences results in performance degradation. 
This motivates automated placement of fences.  Tools that restore sequential
consistency in the program may insert more fences than necessary for the
program to be correct.  Therefore, we propose a property-driven technique
that introduces \emph{reorder-bounded exploration} to identify the smallest
number of program locations for fence placement.  We implemented our
technique on top of \cbmc; however, in principle, our technique is generic
enough to be used with any model checker. Our experimental results show that 
our technique is faster and solves more instances of relevant benchmarks
than earlier approaches.
\end{abstract}

\begingroup
   \renewcommand\thefootnote{}
   \footnotetext{\hspace{-0.85em}$\ddagger$ The published version of this article is available at : 
\url{http://dx.doi.org/10.1007/978-3-319-19249-9_19}
	 }
	 \endgroup
	 
\section{Introduction}\label{Se:intro}

Modern multicore CPUs implement optimizations such as \IT{store buffers} and
\IT{invalidate queues}.  These features result in weaker memory consistency
guarantees than sequential consistency (SC)~\cite{lamportsc}.  Though such
hardware optimizations offer better performance, the weaker consistency has
the drawback of intricate and subtle semantics, thus making it harder for
programmers to anticipate how their program might behave when run on such
architectures.  For example, it is possible for a pair of statements to
appear to have executed out of the program order.

Consider the program given in \figref{tsoreorder}.  Here, \TT{x} and \TT{y}
are shared variables whereas \TT{r1} and \TT{r2} are thread-local variables. 
Statements $s_1$ and $s_3$ perform write operations.  Owing to store
buffering, these writes may not be reflected immediately in the memory. 
Next, both threads may proceed to perform the read operations $s_2$ and
$s_4$.  Since the write operations might still not have hit the memory,
stale values for \TT{x} and \TT{y} may be read in \TT{r2} and \TT{r1},
respectively.  This will cause the assertion to fail.  Such behaviour is
possible with architectures that implement \IT{Total Store Order (TSO)},
which allows write-read reordering.  Note that on a hypothetical
architecture that guarantees sequential consistency, this would never
happen.  However, owing to store buffering, a global observer might witness
that the statements are executed in the order $(s_2,s_4,s_1,s_3)$, which
results in the assertion failure.  We say that $\inroundb{s_1,s_2}$ and
$\inroundb{s_3,s_4}$ have been reordered.

\figref{psoreorder} shows how the assertion might fail on architectures that
implement \IT{Partial Store Order (PSO)}, which permits write-write and
write-read reordering.  Using SC, one would expect to observe $\MT{r2==1}$
if $\MT{r1==1}$ has been observed.  However, reordering of the write
operations $(s_1,s_2)$ leads to the assertion failure.  Architectures such
as Alpha, POWER and SPARC RMO even allow read-write and read-read
reorderings, amongst other behaviours.  Fortunately, all modern
architectures provide various kinds of \IT{memory barriers (fences)} to
prohibit unwanted weakening.  Due to the intricate semantics of weak memory
models and fences, an automated approach to the placement of fences is
desirable.

In this paper, we make the following contributions:
\begin{itemize}
\item We introduce \IT{ReOrder Bounded Model Checking (ROBMC)}. In ROBMC,
the model checker is restricted to exploring only those behaviours of a
program that contain at most $k$ reorderings for a given bound $k$.  The
reorder bound is a new parameter for bounding model checking that has not
been explored earlier.
\item We study how the performance of the analysis is affected as the bound
changes.
\item We implement two ROBMC-based algorithms. In addition, we implement
earlier approaches in the same framework to enable comparison with
ROBMC.
\end{itemize}

The rest of the paper is organized as follows. \sectref{overview} provides
an overview and a motivating example for ROBMC. 
\sectrefs{preliminaries}{wmmrepair} provide preliminaries and describe
earlier approaches respectively.  ROBMC is described in \sectref{reorder}. 
Related research is discussed in \sectref{related}.  Experimental results
are given in \sectref{results}.  Finally, we make concluding remarks in
\sectref{conclusion}.

\section{Motivation and Overview} \label{Se:overview}

\begin{figure}[t]
\begin{center}
\begin{tabular}{c@{\qquad}c@{\qquad}c}

	\subcaptionbox{\label{Fi:tsoreorder}}[.28\linewidth]
	{
\centering
\begin{scriptsize}
\begin{tabular}{ccc}
\multicolumn{3}{c}{$\MT{x=0,y=0;}$} \\
& & \\
\begin{minipage}{0.55in}
$s_1$ :  \MT{x=1;} \\
$s_2$  :  \MT{r1=y;} \\
\end{minipage} & \large{$\parallel$} &
\begin{minipage}{0.55in}
$s_3$  :  \MT{y=1;} \\
$s_4$  :  \MT{r2=x;} \\
\end{minipage} \\
& & \\
\multicolumn{3}{c}{\MT{assert(r1==1||r2==1);}}\\
\end{tabular}
\end{scriptsize}
}
&

\subcaptionbox{\label{Fi:psoreorder}}[.29\linewidth]
{
\centering
\begin{scriptsize}
\begin{tabular}{ccc}
\multicolumn{3}{c}{$\MT{x=0,y=0;}$} \\
& & \\
\begin{minipage}{0.5in}
$s_1$ :  \MT{x=1;} \\
$s_2$  :  \MT{y=1;} \\
\end{minipage} & \large{$\parallel$} &
\begin{minipage}{0.55in}
$s_3$  :  \MT{r1=y;} \\
$s_4$  :  \MT{r2=x;} \\
\end{minipage} \\
& & \\
\multicolumn{3}{c}{\MT{assert(r1!=1||r2==1);}}\\
\end{tabular}
\end{scriptsize}
}
&
\subcaptionbox{\label{Fi:innocent}}[.25\linewidth]
{
\begin{scriptsize}
\begin{tabular}{ccc}
\multicolumn{3}{c}{$\MT{x=0,y=0,w=0,z=0;}$} \\
& & \\
\begin{minipage}{0.55in}
\centering
$s_1$  :  \MT{z=1;} \\
$s_2$  :  \MT{p1=w;} \\
$s_3$ :  \MT{x=1;} \\
$s_4$  :  \MT{r1=y;} \\
\end{minipage} & \large{$\parallel$} &
\begin{minipage}{0.55in}
$s_5$  :  \MT{w=1;} \\
$s_6$  :  \MT{p2=z;} \\
$s_7$  :  \MT{y=1;} \\
$s_8$  :  \MT{r2=x;} \\
\end{minipage} \\
& & \\
\multicolumn{3}{c}{\MT{assert(r1==1||r2==1);}} \\
\multicolumn{3}{c}{ \MT{assert(p1+p2>=0);}}\\
\end{tabular}
\end{scriptsize}
}
\end{tabular}
\end{center}
\caption{(\subref{Fi:tsoreorder}) Reordering in TSO. (\subref{Fi:psoreorder}) Reordering in PSO.
(\subref{Fi:innocent}) A program with \IT{innocent} and \IT{culprit} reorderings }
\end{figure}

There has been a substantial amount of previous research on automated fence
insertion~\cite{remmex-pso,memorax-tool,dfence,trencher,jadefmsd12,fender,cav2014}. 
We distinguish approaches that aim to restore sequential consistency (SC)
and approaches that aim to ensure that a user-provided assertion holds. 
Since every fence incurs a performance penalty, it is desirable to keep the
number of fences to a minimum.  Therefore, a property-driven approach for
fence insertion can result in better performance.  The downside of the
property-driven approach is that it requires an explicit specification.

Consider the example given in \figref{innocent}. Here, \TT{x,y,z,w} are
shared variables initialized to $0$.  All other variables are thread-local. 
A processor that implements total store ordering (TSO) permits a read of a
global variable to precede a write to a different global variable when there
are no dependencies between the two statements.  Note that if $(s_3,s_4)$ or
$(s_7,s_8)$ is reordered, the assertion will be violated.  We shall call
such pairs of statements \IT{culprit pairs}. By contrast, the pairs
$(s_1,s_2)$ and $(s_5,s_6)$ do not lead to an assertion violation
irrespective of the order in which their statements execute.  We shall
call such pairs \IT{innocent pairs}.  A tool that restores SC would insert
four fences, one for each pair mentioned earlier.  However, only two fences
(between $s_3,s_4$ and $s_7,s_8$) are necessary to avoid the assertion
violation.

Some of the earlier property-driven techniques for fence
insertion~\cite{remmex-tso,memorax-tool} use the following approach. 
Consider a counterexample to the assertion.  Every counterexample to the
assertion must contain at least one culprit reordering.  If we prevent all
culprit reorderings, the program will satisfy the property.  This is done in
an iterative fashion.  For all the counterexamples seen, a smallest set of
reorderings $S$ is selected such that $S$ has at least one reordering in
common with each of the counterexamples.  Let us call such a set a
\IT{minimum-hitting-set} ($\mathit{MHS}$) over all the set of
counterexamples $C$ witnessed so far.  All the weakenings in $\mathit{MHS}$
are excluded from the program.  Even though $\mathit{MHS}$ may not cover all
the culprit reorderings initially, it will eventually consist of culprit
pairs only.  Since one cannot distinguish the innocent pairs from the
culprit ones a priori, such an approach may get distracted by innocent
pairs, thus, taking too long to identify the culprit pairs.

To illustrate, let us revisit the example in \figref{innocent}. Let us name
the approach described above \FI (Fence Insertion).  Let the first
counterexample path $\pi^1$ be $(s_2,s_1,s_6,s_5,s_4,s_7,s_8,s_3)$.  The set
of reorderings is $\inparan{(s_1,s_2),(s_3,s_4),(s_5,s_6)}$.  Method \FI may
choose to forbid the reordering of $\inparan{(s_1,s_2)}$, as it is one of
the choices for the $\mathit{MHS}$.  Next, let
$\pi^2=(s_1,s_2,s_6,s_5,s_4,s_7,\allowbreak s_8,s_3)$.  The set of
reorderings for this trace is $\inparan{(s_3,s_4),(s_5,s_6)}$.  There are
multiple possible choices for $\mathit{MHS}$.  For instance, \FI may choose to forbid
$\inparan{(s_5,s_6)}$.  Let $\pi^3=(s_2,s_1,s_5,s_6,s_8,s_3,s_4,s_7)$.  As
the set of reorderings is $\inparan{(s_1,s_2),(s_7,s_8)}$, one of the
choices for the $\mathit{MHS}$ is $\inparan{(s_1,s_2), (s_5,s_6)}$.  Recall that
$(s_1,s_2)$ and $(s_5,s_6)$ are innocent pairs.  On the other hand,
$(s_3,s_4)$ and $(s_7,s_8)$ are culprit pairs.  \FI may continue with
$\pi^4=(s_1,s_2,s_5,s_6,s_4,s_7,s_8,s_3)$.  The set of reorderings in
$\pi^4$ is $\inparan{(s_3,s_4)}$.  An adversarial $\mathit{MHS}$ would be
$\inparan{(s_1,s_2),(s_3,s_4)}$.  Let $\pi^5$ be
$(s_1,s_2,s_6,s_5,s_8,\allowbreak s_3,s_4,s_7)$.  The reorderings
$\inparan{(s_5,s_6),(s_7,s_8)}$ will finally lead to the solution
$\inparan{(s_3,s_4),(s_7,s_8)}$.  In the $6^{\mbox{\scriptsize th}}$
iteration \FI will find that the program is safe with a given $\mathit{MHS}$.  For
brevity, we have not considered traces with reorderings $(s_1,s_4)$ and
$(s_5,s_8)$.  In the worst case, considering these reorderings might lead to
even more traces. 

As we can see, the presence of innocent pairs plays a major role in how fast
\FI will be able to find the culprit pairs.  Consider a program with many
more innocent pairs.  \FI will require increasingly more queries to the
underlying model checker as the number of innocent pairs increases.

To address the problem caused by innocent pairs, we propose \IT{Reorder
Bounded Model Checking} (\robmc).  In \robmc, we restrict the model checker
to exploring only the behaviours of the program that have at most $k$
reorderings for a given reordering bound $k$.  Let us revisit the example
given in \figref{innocent} to see how the bounded exploration affects the
performance.  Assume that we start with the bound $k=1$.  Since the
model checker is forced to find a counterexample with only one reordering,
there is no further scope for an innocent reordering to appear in the
counterexample path.  Let the first trace found be
$\pi^1=(s_1,s_2,s_4,s_5,s_6,s_7,s_8,s_3)$.  There is only one reordering
$\inparan{(s_3,s_4)}$ in this trace.  The resulting $\mathit{MHS}$ will be
$\inparan{(s_3, s_4)}$.  Let the second trace be
$\pi^2=(s_1,s_2,s_5,s_6,s_8,s_3,s_4,s_7)$.  As the only reordering is
$\inparan{(s_7,s_8)}$, the $\mathit{MHS}$ over these two traces would be
$\inparan{(s_3,s_4)(s_7,s_8)}$.  The next query would declare the program
safe.  Now, even with a larger bound, no further counterexamples can be
produced.  This example shows how a solution can be found much faster with
\robmc compared to \FI.  In the following sections, we describe our approach
more formally.

\section{Preliminaries} \label{Se:preliminaries}

Let $P$ be a concurrent program.  A program execution is a sequence of
events.  An event $e$ is a four-tuple
\[ e \equiv \tuple{\mathit{tid},\mathit{in},\mathit{var},\mathit{type}} \]
where $tid$ denotes the thread identifier associated with the event and $in$
denotes the instruction that triggered the event.  Instructions are dynamic
instances of program statements.  A program statement can give rise to
multiple instructions due to loops and procedure calls.  $stmt : Instr
\rightarrow Stmt$ denotes a map from instructions to their corresponding
program statements.  The program order between any two instructions $I_1$ and
$I_2$ is denoted as $I_1 \PO I_2$, which indicates that $I_1$ precedes $I_2$
in the program order.  The component $var$ denotes the global/shared
variable that participated in the event $e$.  The type of the event is
represented by $\mathit{type}$, which can either be $\mathit{read}$ or
$\mathit{write}$.  Without loss of generality, we assume that $P$ only
accesses one global/shared variable per statement.  Therefore, given a
statement $s \in Stmt$, we can uniquely identify the global variable
involved as well as the type of the event that $s$ gives rise to.  Any
execution of program $P$ is a sequence of events $\pi = (e_1,\dots,e_n)$.
%
%
The $i^{\mbox{\scriptsize th}}$ event in the sequence $\pi$ is denoted by~$\pi(i)$. 
\begin{mydefinition}\label{De:reorder}
A pair of statements $(s_1,s_2)$ of a program is said to be \EM{reordered}
in an execution $\pi$ if:
\begin{align*}
	\exists _i \exists _j \left(  \inroundb{e_i.tid=e_j.tid} \wedge \inroundb{\pi(i) =
	e_i} \right.
	\wedge \inroundb{\pi(j)=e_j} \\ \wedge \inroundb{j < i}  \wedge
	\inroundb{e_i.in=I_1 \wedge e_j.in=I_2} \\ \left. \wedge \inroundb{I_1 \PO I_2} 
\wedge \inroundb{stmt(I_1)=s_1 \wedge stmt(I_2)= s_2} \right)
\end{align*}
\end{mydefinition}

According to \defref{reorder}, two statements are reordered if they give
rise to events that occurred out of program order.
\begin{mydefinition}
We write $RO_A(s_1,s_2)$ to denote that an architecture $A$ allows
the pair of statements $(s_1,s_2)$ to be reordered.
\end{mydefinition}

Different weak memory architectures permit particular reorderings of events.
\begin{itemize}
\item \BF{Total Store Order (TSO)}: TSO allows a read to be reordered before
 a write if they access different global variables.
 \begin{align*}
  RO_{tso}(s_1,s_2)  \equiv &  \inroundb{s_1.var \neq s_2.var}   \wedge \inroundb{s_1.type=write \wedge s_2.type=read}
\end{align*}
\item \BF{Partial Store Order (PSO)}: PSO allows a read or write to be reordered
 before a write if they access different global variables.
 \begin{align*}
  RO_{pso}(s_1,s_2)  \equiv &  \inroundb{s_1.var \neq s_2.var}   \wedge \inroundb{s_1.type=write}
\end{align*}
\end{itemize}

Partial-order based models for TSO, PSO, \IT{read memory order (RMO)} and
POWER are presented in detail in~\cite{jadefmsd12}.

\begin{mydefinition}
Let $C$ be a set consisting of non-empty sets $S_1,\dots,S_n$.
The set \MC{H} is called a \EM{hitting-set} (HS) of $C$ if: 
\[ \forall_{S_i \in C} \MC{H} \cap S_i \neq \emptyset \]
\MC{H} is called a \EM{minimal-hitting-set} (mhs) if any proper subset of \MC{H}
is not a hitting-set.  \MC{H} is a \EM{minimum-hitting-set} (MHS) of $C$ if $C$
does not have a smaller hitting-set.  Note that a collection~$C$ may have
multiple minimum-hitting-sets.
\end{mydefinition}

\section{Property-driven Fence Insertion}\label{Se:wmmrepair}

\subsection{Overview}

In this section we will discuss two approaches that were used
earlier for property-driven fence insertion.  We will present our
improvements in the next section.

For a program $P$ of size $\vbars{P}$, the total number of pairs of
statements is $\vbars{P}^2$.  Since the goal is to find a subset of these
pairs, the search space is $2^{\vbars{P}^2}$.  Thus, the search space grows
\EM{exponentially} as the size of the program is increased.

An automated method for fence insertion typically includes two components:
(1)~a model checker $M$ and (2)~a search technique that uses $M$ iteratively
in order to find a solution.  We assume that the model checker $M$ has the
following properties:
\begin{itemize}
\item $M$ should be able to find counterexamples to assertions
in programs given a memory model.
\item $M$ should return the counterexample $\pi$ in form of a sequence of
events as described in \sectref{preliminaries}.
\item For a pair of statements $(s_1,s_2)$ for which $RO_A(s_1,s_2)$ holds,
$M$ should be able to enforce an ordering constraint $s_1 \prec s_2$ that
forbids the exploration of any execution where $(s_1,s_2)$ is reordered.
\end{itemize}

\subsection{Fence Insertion using Trace Enumeration}

\algref{te} is a very simple approach to placing fences in the program with
the help of such a model checker.  The algorithm is representative of the
technique that is used in \dfence~\cite{dfence}.  \algref{te} iteratively
submits queries to $M$ for a counterexample \algline{algte:mcall}.  All the
pairs of statements that have been reordered in $\pi$ are collected in $SP$
\algline{algte:getpair}.  To avoid the same trace in future iterations,
reordering of at least one of these pairs must be disallowed.  The choice of
which reorderings must be banned is left open.  This process is repeated
until no further error traces are found.  Finally,
$\mathit{computeMinimalSolution}(\phi)$ computes a minimal set of pairs of
statements such that imposing ordering constraints on them satisfies~$\phi$.

\paragraph{\bf Termination and soundness}

Even though the program may have unbounded loops and thus potentially
contains an unbounded number of counterexamples, \algref{te} terminates. 
The reason is that an ordering constraint $s_1 \prec s_2$ disallows
reordering of all events that are generated by $(s_1,s_2)$.  The number of
iterations is bounded above by $2^{\vbars{P}^2}$, which is the size of the
search space.  Soundness is a consequence of the fact that the algorithm
terminates only when no counterexamples are found.  A minimal-hitting-set
(mhs) is computed over all these counterexamples to compute the culprit
pairs that must not be reordered.  Since every trace must go through one of
these pairs, it cannot manifest when the reordering of these pairs is
banned.  The number of pairs computed is minimal, thus, \algref{te} does not
guarantee the least number of fences.  One can replace the
minimal-hitting-set (mhs) with a minimum-hitting-set (MHS) in order to
obtain such a guarantee.

\begin{algorithm}[pt]
\caption{Trace Enumerating Fence Insertion (\TE)}
\label{Alg:te}
\begin{scriptsize}
\begin{algorithmic}[1]
\STATE \BF{Input: } Program $P$
\STATE \BF{Output: } Set $S$ of pairs of statements that must not be reordered to avoid
assertion failure
\STATE $C:=\emptyset$
\STATE $S:= \emptyset$
\STATE $\phi:= \mathit{true}$ \label{algte:constraint}
\LOOP
\STATE $\inangleb{\mathit{result},\pi}:=M(P_\phi)$ \label{algte:mcall}
\IF {$\mathit{result}=\mathit{SAFE}$} \label{algte:safestart}
\STATE \BF{break}
\ENDIF \label{algte:safeend}
\STATE $\mathit{SP}:= \mathit{GetReorderedPairs}(\pi)$ \label{algte:getpair}
\IF {$\mathit{SP} = \emptyset$} \label{algte:bcstart}
\PRINT \BF{Error:} Program cannot be repaired
\RETURN \BF{errorcode}
\ENDIF \label{algte:bcend}
\STATE $\phi := \displaystyle \phi \wedge \inroundb{\bigvee _{(s_1,s_2) \in SP} s_1 \prec s_2}$ \label{algte:enforce}
\ENDLOOP
\STATE $S := \mathit{computeMinimalSolution}(\phi)$
\RETURN $S$
\end{algorithmic}
\end{scriptsize}
\end{algorithm}

\begin{algorithm}[pt]
\caption{Accelerated Fence Insertion (\FI)}
\label{Alg:wmmprogrepair}
\begin{scriptsize}
\begin{algorithmic}[1]
\STATE \BF{Input: } Program $P$
\STATE \BF{Output: } Set $S$ of pairs of statements that must not be reordered to avoid
assertion failure
\STATE $C:=\emptyset$
\STATE $S:= \emptyset$
\STATE $\phi:= \mathit{true}$ \label{algrep:constraint}
\LOOP
\STATE $\inangleb{\mathit{result},\pi}:=M(P_\phi)$ \label{algrep:mcall}
\IF {$\mathit{result}=\mathit{SAFE}$} \label{algrep:safestart}
\STATE \BF{break}
\ENDIF \label{algrep:safeend}
\STATE $\mathit{SP} := \mathit{GetReorderedPairs}(\pi)$ \label{algrep:getpair}
\IF {$\mathit{SP} = \emptyset$} \label{algrep:bcstart}
\PRINT \BF{Error:} Program cannot be repaired
\RETURN \BF{errorcode}
\ENDIF \label{algrep:bcend}
\STATE \highlight{$C := C \cup \inparan{\mathit{SP}}$}
\STATE \highlight{$S:=\mathit{MHS}(C)$ \label{algrep:mhs}}
\STATE \highlight{$\phi := \displaystyle \bigwedge _{(s_1,s_2) \in S} s_1 \prec s_2$} \label{algrep:enforce}
\ENDLOOP
\RETURN $S$
\end{algorithmic}
\end{scriptsize}
\end{algorithm}

\subsection{Accelerated Fence Insertion}

\algref{wmmprogrepair} is an alternative approach to fence insertion.  The
differences between \algref{te} and \algref{wmmprogrepair} are highlighted. 
\algref{wmmprogrepair} has been used in~\cite{remmex-tso,remmex-pso} and is
a variant of the approach used in~\cite{memorax-tool}. 
\algref{wmmprogrepair} starts with an ordering constraint $\phi$
\algline{algrep:constraint}, which is initially unrestricted.  A call to the
model checker $M$ is made \algline{algrep:mcall} to check whether the
program $P$ under the constraint $\phi$ has a counterexample.  From a
counterexample~$\pi$, we collect the set of pairs of statements $SP$ that
have been reordered in $\pi$ \algline{algrep:getpair}.  This set is put into
a collection~$C$.

Next, we compute a minimum-hitting-set over~$C$.  This gives us one of the
smallest sets of pairs of statements that can avoid all the counterexamples
seen so far.  The original approach in~\cite{remmex-tso} uses a
minimal-hitting-set (mhs).  The ordering constraint~$\phi$ is updated using
the minimum-hitting-set \alglines{algrep:mhs}{algrep:enforce}. 
\algref{wmmprogrepair} tells the model checker which reorderings from each
counterexample are to be banned at every iteration, which is in contrast to
\algref{te}.  \algref{wmmprogrepair} assumes that an assertion violation in
$P$ is due to a reordering.  If a counterexample is found without any
reordering, the algorithm exits with an
error~\alglines{algrep:bcstart}{algrep:bcend}.  Finally, the algorithm
terminates when no more counterexamples can be found
\alglines{algrep:safestart}{algrep:safeend}.

\paragraph{\bf Termination and soundness}

The argument that applies to \algref{te} can also be used to prove
termination and soundness of \algref{wmmprogrepair}.  In addition, the
constraint $\phi$ generated is generally stronger
(i.e.~$\phi_\textrm{\algref{wmmprogrepair}} \rightarrow \phi_\textrm{\algref{te}}$) than the
constraint generated by \algref{te}.  Thus, for the same sequence of traces,
\algref{wmmprogrepair} typically converges to a solution faster than
\algref{te}.

\section{Reorder-bounded Exploration} \label{Se:reorder}

\algref{wmmprogrepair} can further be improved by avoiding innocent
reorderings so that culprit reorderings responsible for the violation
of the assertion are found faster.

As discussed in \sectref{overview},  \algref{wmmprogrepair} requires many
iterations to converge and terminate in the presence of innocent
reorderings.  The reason is that the model checker may not return the
simplest possible counterexample that explains the assertion violation due
to reorderings.  In order to address this problem, we need a model checker
$M'$ with an additional property as follows:
\begin{itemize}
\item $M'$ takes $P_{\phi}$ and $k$ as inputs. Here, $P_{\phi}$ is the program
	along with the ordering constraint~$\phi$ and $k$ is a positive integer.
	$M'$ produces a counterexample $\pi$ for $P_{\phi}$ such that $\pi$ has 
	at most $k$ reorderings. If it cannot find a counterexample with at most
	$k$ reorderings, then it will declare $P_{\phi}$ safe.
\end{itemize}

With a model checker $M'$, we can employ  \algref{brepair} to speed up the
discovery of the smallest set of culprit pairs of statements.  The
steps that differ from \algref{wmmprogrepair} in \algref{brepair} are
highlighted.  \algref{brepair} initializes the reordering bound $k$
\algline{algbrep:binit} to a given lower bound~$K_1$.  The model checker $M'$
is now called with this bound to obtain a counterexample that has at most
$k$ reorderings \algline{algbrep:mcall}.  When the counterexample cannot be
found, the bound~$k$ is increased according to some strategy denoted by
$\mathit{increaseStrategy}$ \algline{algbrep:increase}.  Note that
collection~$C$ and the ordering constraint $\phi$ are preserved even when
$k$ is increased.  Thus, when $k$ is increased from $k_1$ to $k_2$, the
search for culprit reorderings starts directly with the ordering
constraints that repair the program for up to $k_1$ reorderings.  Only those
counterexamples that require more than $k_1$ and fewer than $k_2$
culprit reorderings will be reported.  Let us assume that $P$ does not
have any counterexample with more than $k_{opt}$ reorderings.  If $k_{opt}$
is much smaller than $k$, the performance of \algref{brepair} might suffer
due to interference from innocent reorderings.  If the increase in $k$
is too small, the algorithm might have to go through many queries to reach
the given upper bound $K_2$.  It can be beneficial to increase the bound $k$
by a larger amount after witnessing a few successive $\MI{SAFE}$ queries,
and by a smaller amount when a counterexample has been found recently.

\henabledfalse
\renewcommand{\algorithmicendwhile} {\ifhenabled \highlight{\algorithmicend\
\algorithmicwhile} \else \algorithmicend\ \algorithmicwhile \fi}
\renewcommand{\algorithmicwhile}{ \ifhenabled \highlight{\BF{while}} \else \BF{while} \fi }
\renewcommand{\algorithmicdo}{\ifhenabled \highlight{\BF{do}} \else \BF{do} \fi }

\begin{algorithm}[t]
\caption{\SC{ROBMC}}
\label{Alg:brepair}
\begin{scriptsize}
\begin{algorithmic}[1]
\STATE \BF{Input: } Program $P$, lower bound $K_1$ and an upper bound $K_2$
\STATE \BF{Output: } Set $S$ of pairs of statements that must not be reordered to avoid
assertion failure
\STATE $C:=\emptyset$
\STATE $S := \emptyset$
\STATE \highlight{$k := K_1$} \label{algbrep:binit}
\STATE $\phi := \mathit{true}$ \label{algbrep:constraint} \henabledtrue
\WHILE {\highlight{ $k\leq K_2$} }
\LOOP
\STATE \highlight{$\inangleb{\mathit{result},\pi}:=M'(P_\phi,k)$} \label{algbrep:mcall}
\IF {$\mathit{result}=\mathit{SAFE}$} \label{algbrep:safestart}
\STATE \BF{break}
\ENDIF \label{algbrep:safeend}
\STATE $\mathit{SP} := \mathit{GetReorderedPairs}(\pi)$ \label{algbrep:getpair}
\IF {$\mathit{SP} = \emptyset$} \label{algbrep:bcstart}
\PRINT \BF{Error:} Program cannot be repaired
\RETURN \BF{errorcode}
\ENDIF \label{algbrep:bcend}
\STATE $C := C \cup \inparan{SP}$
\STATE $S:=\mathit{MHS}(C)$ \label{algbrep:mhs}
\STATE $\phi := \displaystyle \bigwedge _{(s_1,s_2) \in S} s_1 \prec s_2$ \label{algbrep:enforce}
\ENDLOOP
\STATE \highlight{$k:=\mathit{increaseStrategy}(k)$} \label{algbrep:increase}
\ENDWHILE
\RETURN $S$

\end{algorithmic}
\end{scriptsize}
\end{algorithm}

\renewcommand{\algorithmicif}{\ifhenabled \highlight{\BF{if}} \else \BF{if} \fi }
\renewcommand{\algorithmicthen}{\ifhenabled \highlight{\BF{then}} \else \BF{then} \fi }
\renewcommand{\algorithmicendif} {\ifhenabled \highlight{\algorithmicend\
\algorithmicif} \else \algorithmicend\ \algorithmicif \fi}

\begin{algorithm}[t]
\caption{\SC{ROBMC-Et}}
\label{Alg:obrepair}
\begin{scriptsize}
\begin{algorithmic}[1]
\STATE \BF{Input: } Program $P$, lower bound $K_1$ and an upper bound $K_2$
\STATE \BF{Output: } Set $S$ of pairs of statements that must not be reordered to avoid
assertion failure
\STATE $C:=\emptyset$
\STATE $S := \emptyset$
\STATE $k := K_1$ \label{algobrep:binit}
\STATE $\phi := \mathit{true}$ \label{algobrep:constraint} 
\STATE \highlight{$\mathit{terminate} := \mathit{false}$} 
\WHILE {  $k\leq K_2$  \AND \highlight{$\mathit{terminate}=\mathit{false}$}}
\LOOP
\STATE \highlight{$\inangleb{\mathit{result},\pi,\psi}:=M'(P_\phi,k)$} \label{algobrep:mcall}
\henabledfalse
\IF {$\mathit{result}=\mathit{SAFE}$} \label{algobrep:safestart}  \henabledtrue
\IF{ \highlight{ \NOT $\mathit{safeDueToBound}(k,\psi)$} } \label{algobrep:etstart}
\STATE \highlight{$\mathit{terminate}:=\mathit{true}$} \label{algobrep:et}
\ENDIF \label{algobrep:etend}
\STATE \BF{break}
\ENDIF \label{algobrep:safeend}
\STATE $SP := \mathit{GetReorderedPairs}(\pi)$ \label{algobrep:getpair}
\IF {$SP = \emptyset$} \label{algobrep:bcstart}
\PRINT \BF{Error:} Program cannot be repaired
\RETURN \BF{errorcode}
\ENDIF \label{algobrep:bcend}
\STATE $C := C \cup \inparan{SP}$
\STATE $S:=\mathit{MHS}(C)$ \label{algobrep:mhs}
\STATE $\phi := \displaystyle \bigwedge _{(s_1,s_2) \in S} s_1 \prec s_2$ \label{algobrep:enforce}
\ENDLOOP
\STATE $k:=\mathit{increaseStrategy}(k)$ \label{algobrep:increase}
\ENDWHILE
\RETURN $S$
\end{algorithmic}
\end{scriptsize}
\end{algorithm}

\paragraph{\bf Building $M'$} 

A model checker $M'$ that supports bounded exploration can be constructed
from $M$ as follows.  For every pair $(s_1,s_2)$ that can potentially be
reordered, we introduce a new auxiliary Boolean variable $a_{12}$.  Then, a
constraint $\neg a_{12} \leftrightarrow (s_1 \prec s_2)$ can be added.  This
allows us to enforce the ordering constraint $s_1 \prec s_2$ by manipulating
values assigned to $a_{12}$.  For a given bound $k$, we can enforce a
reorder-bounded exploration by adding a cardinality constraint
$\displaystyle \sum a_{ij} \leq k$.  This constraint forces only up to $k$
auxiliary variables to be set to $\mathit{true}$, thus, allowing only up to
$k$ reorderings.

\paragraph{\bf Optimizing \algref{brepair}}

Even when the correct solution for the program is found, \algref{brepair}
has to reach the upper bound $K_2$ to terminate.  This can cause many
further queries for which the model checker $M'$ is going to declare the
program $\MI{SAFE}$.  To achieve soundness with \algref{brepair}, $K_2$
should be as high as the total number of all the pairs of statements that
can be potentially reordered.  This leads to a very high value for $K_2$,
which may reduce the advantage that \algref{brepair} has over
\algref{wmmprogrepair}.

We can avoid these unnecessary queries if the model checker $M'$ produces a
proof whenever it declares the program $P_{\phi}$ as $\MI{SAFE}$.  This
proof is analogous to an \IT{unsatisfiable core} produced by many SAT/SMT
solvers whenever the result of a query is \IT{unsat}.\footnote{SAT solvers
such as MiniSat~\cite{minisat} and Lingeling~\cite{lingeling} allow to query
whether a given assumption was part of the unsatisfiable
core~\cite{DBLP:journals/entcs/EenS03}.} With this additional feature of
$M'$, we can check whether the cardinality constraint $\sum a_{ij} \leq k$
was the reason for declaring the program $\MI{SAFE}$.  If not, we know that
$P$ is safe under the ordering constraint~$\phi$ irrespective of the bound. 
Therefore, \algref{brepair} can terminate early as shown in
\algref{obrepair}.  The difference between \algref{brepair} and
\algref{obrepair} is highlighted in \algref{obrepair}.  The model checker
$M'$ now returns $\psi$ as a proof when $P_{\phi}$ is safe
\algline{algobrep:mcall}.  When $M'$ declares $P_\phi$ as safe,
\algref{obrepair} checks whether the bound $k$ is the reason that $P_\phi$
is declared safe \algline{algobrep:etstart}.  If not, the termination flag
is set to $\MI{true}$ to trigger early termination \algline{algobrep:et}.

\paragraph{\bf Termination and soundness}

Let the program $P$ have counterexamples with up to $k_{opt}$ culprit
reorderings.  If the value of the upper bound $K_2$ for \algref{brepair} and
\algref{obrepair} is smaller than $k_{opt}$, there might exist traces that
the algorithms fail to explore.  For soundness, the value of $K_2$ should
thus be higher than $k_{opt}$.  Since $k_{opt}$ is generally not known a
priori, a conservative value of $K_2$ should be equal to the total number of
pairs of statements for which reordering might happen ($RO_A(s_1,s_2)$ is
\MI{true}).  Termination is guaranteed due to finiteness of the number of
pairs of statements and~$K_2$.

\section{Related work} \label{Se:related}

There are two principal approaches for modelling weak memory semantics.  One
approach is to use operational models that explicitly model the buffers and
queues to mimic the
hardware~\cite{remmex-pso,vechevpowmm,dfence,memorax-paper,DBLP:conf/esop/AlglaveKNT13,trencher,NETYS2015}. 
The other approach is to axiomatize the observable behaviours using partial
orders~\cite{wmmcav13,jadefmsd12,wmmcpp}.  Buffer-based modelling is closer
to the hardware implementation than the partial-order based approach. 
However, the partial-order based approach provides an abstraction of the
underlying complexity of the hardware and has been proven
effective~\cite{wmmcav13}.  Results on complexity and decidability for
various weak memory models such as TSO, PSO and RMO are given
in~\cite{sebastian-wmm}.

Due to the intricate and subtle semantics of weak memory consistency and the
fences offered by modern architectures, there have been numerous efforts
aimed at automating fence
insertion~\cite{memorax-tool,jadefmsd12,trencher,pensieve,fender,remmex-tso,remmex-pso,dfence,cav2014}. 
These works can be divided into two categories.  In one category, fences are
inserted in order to restore sequential
consistency~\cite{jadefmsd12,trencher,cav2014}.  The primary advantage is
that no external specification is required.  On the downside, the fences
inferred by these methods may be unnecessary.

The second category are methods that insert only those fences that are
required for a program to satisfy given
properties~\cite{remmex-tso,remmex-pso,memorax-tool,dfence,NETYS2015}. 
These techniques usually require repetitive calls to a model checker or a
solver.  \dfence is a dynamic analysis tool that falls into this category. 
Our work differs from \dfence as ours is a fully static approach as compared
to the dynamic approach used by \dfence.  A direct comparison with \dfence
cannot be made. However, we have implemented their approach in our framework
and we present an experimental comparison using our re-implementation.

\memorax~\cite{memorax-tool} and \remmex~\cite{remmex-tso,remmex-pso} also
fall into the category of property-driven tools. 
\memorax~\cite{memorax-paper} computes all possible minimal-hitting-set
solutions.  Though it computes the smallest possible solution, exhaustively
searching for all possible solutions can make such an approach slow. 
Moreover, \memorax requires that the input program is written in \SC{rmm} |
a special purpose language.  \algref{wmmprogrepair} captures what \memorax
would do if it has to find only one solution.  \remmex also falls in the
category of property-driven tools and their approach is given as
\algref{wmmprogrepair}.

Bounded model checking has been used for the verification of concurrent
programs~\cite{wmmcav13,contextbmc}.  In context-bounded model
checking~\cite{contextbmc, boolcontextbmc}, the number of interleavings in
counterexamples is bounded, but executions are explored without depth limit. 
ROBMC is orthogonal to these ideas, as here the bound is on the number of
event reorderings.

\section{Implementation and Experimental Results}\label{Se:results}

\begin{figure}[t]
\centering
\begin{tabular}{ccc}
\multicolumn{3}{c} {$\MT{[x_i=0;\,y_i=0;]}^{n} $} \\
\multicolumn{3}{c} {$ \MT{s1=0;\,s2=0;}$} \\
& & \\
\begin{minipage}[t]{0.25\textwidth}
	\centering
$\insquareb{
\begin{array}{c}
\MT{x_i=1;} \\
\MT{s1\mbox{+=}y_i;} \\
\end{array}
} ^{n} $ \\
 
\end{minipage} & \large{$\displaystyle \parallel$} &
\begin{minipage}[t]{0.25\textwidth}
	\centering
$ \insquareb{
\begin{array}{c}
\MT{y_i=1;} \\
\MT{s2\mbox{+=}x_i;} \\
\end{array}
} ^{n} $ \\
 
\end{minipage} \\
& & \\
\multicolumn{3}{c}{$\MT{assert(s1+s2>=0);}$} \\
\end{tabular}
\caption {A parameterized program. 
Here, $[\texttt{st}]^n$ denotes that the statement \texttt{st} is repeated  $n$ times.}
\label{Fi:paramsize}
\end{figure}

\begin{figure}[!tp]
\subcaptionbox{\# of instances solved\label{Fi:instances}}[.5\linewidth]
{
	\begin{tikzpicture}[scale=.65]
		\begin{axis} [ ymin=1, ymax=140, xmin=2, xmax=13, xlabel=$K_1$, ylabel=\# instances solved,
				legend entries={\small{tso-te},tso-fi,tso-robmc,tso-robmc-et,pso-te,pso-fi,pso-robmc,pso-robmc-et},
				legend style= { legend pos=south west, }
			]
			\addplot [dashed,domain=2:13] {18};
			\addplot [dashdotted,domain=2:13] {83};
			\addplot  [dashed, mark=x] table {data/num-instances-tso-noet.dat};
			\addplot  [dashdotted,mark=+] table {data/num-instances-tso-et.dat};
			\addplot [mark=triangle,domain=2:13] {4};
			\addplot [mark=diamond,domain=2:13] {46};
			\addplot  [mark=oplus] table {data/num-instances-pso-noet.dat};
			\addplot  [mark=o] table {data/num-instances-pso-et.dat};
		\end{axis}
	\end{tikzpicture}
}
\subcaptionbox{\# of statement pairs \label{Fi:st-pairs}}[.5\linewidth]
{
		\begin{tikzpicture}[scale=.65]
		\begin{axis} [ xmin=2, xmax=40, xlabel=size parameter $n$, ylabel=\# of statement pairs, 
				legend entries={tso,pso},
				legend style= { legend pos=south east, }
			]
			\addplot  [dashed, mark=x] table {data/group-peterson-tso.dat};
			\addplot  [mark=triangle] table {data/group-peterson-pso.dat};
		\end{axis}
	\end{tikzpicture}

}
\subcaptionbox{\robmcet (with $K_1=5$) v/s \FI \label{Fi:scatterk5}}[.5\linewidth]
	{
	\begin{tikzpicture}[scale=.65]
		\begin{loglogaxis} [xmin=1,xmax=600, ymin=1, ymax=600, xlabel=fi,
			ylabel=robmc-et, 
			]
			\addplot [domain=1:600] {x};
			\addplot [mark size=1.5,only marks] table
			{data/scatter-k5-et.dat};
		\end{loglogaxis}
	\end{tikzpicture}
}
\subcaptionbox{peterson on TSO ($K_1=5$) \label{Fi:queries}}[.45\linewidth]
{
	\begin{tikzpicture}[scale=.7]
		\begin{semilogyaxis} [ xmin=2, xmax=40, xlabel=size parameter $n$, ylabel=\#queries,title=peterson-tso,
				legend entries={te,fi,robmc,robmc-et},
				legend style= { at={(.5,.9)},anchor=north}, 
			]
			\addplot  [dashed, mark=x] table {data/query-peterson-tso-te.dat};
			\addplot  [mark=square] table {data/query-peterson-tso-nop.dat};
			\addplot  [mark=oplus] table {data/query-peterson-tso-noet-5.dat};
			\addplot  [mark=triangle] table {data/query-peterson-tso-et-5.dat};
		\end{semilogyaxis}
	\end{tikzpicture}

}

\subcaptionbox{peterson on TSO ($K_1=5$) \label{Fi:peterson-tso-k5}}[.5\linewidth]
{
	\begin{tikzpicture}[scale=.7]
		\begin{axis} [ ymin=0, ymax=600, xmin=2, xmax=40, xlabel=size parameter $n$, ylabel=time(sec),title=peterson-tso,
				legend entries={te,fi,robmc,robmc-et},
				legend style= { at={(.45,.95)},anchor=north }
			]
			\addplot  [dashed, mark=x] table {data/peterson-tso-te.dat};
			\addplot  [mark=square] table {data/peterson-tso-nop.dat};
			\addplot  [mark=oplus] table {data/peterson-tso-noet-k5.dat};
			\addplot  [mark=triangle] table {data/peterson-tso-et-k5.dat};
		\end{axis}
	\end{tikzpicture}

}
\subcaptionbox{peterson on PSO ($K_1=5$)\label{Fi:peterson-pso-k5}}[.5\linewidth]
{
		\begin{tikzpicture}[scale=.7]
		\begin{axis} [ ymin=0, ymax=600, xmin=2, xmax=40, xlabel=size parameter $n$, ylabel=time(sec), title=peterson-pso,
				legend entries={te,fi,robmc,robmc-et},
				legend style= { at={(.45,.95)},anchor=north }
			]
			\addplot  [dashed, mark=x] table {data/peterson-pso-te.dat};
			\addplot  [mark=square] table {data/peterson-pso-nop.dat};
			\addplot  [mark=oplus] table {data/peterson-pso-noet-k5.dat};
			\addplot  [mark=triangle] table {data/peterson-pso-et-k5.dat};
		\end{axis}
	\end{tikzpicture}

}
\phantomcaption
\end{figure}
\begin{figure}
\ContinuedFloat
\subcaptionbox{dijkstra on TSO ($K_1=5$) \label{Fi:dijkstra-tso-k5}}[.5\linewidth]
{
	\begin{tikzpicture}[scale=.7]
		\begin{axis} [ ymin=0, ymax=600, xmin=2, xmax=40, xlabel=size parameter $n$, ylabel=time(sec),title=dijkstra-tso,
				legend entries={te,fi,robmc,robmc-et},
				legend style= { at={(.35,.95)},anchor=north }
			]
			\addplot  [dashed, mark=x] table {data/dijkstra-tso-te.dat};
			\addplot  [mark=square] table {data/dijkstra-tso-nop.dat};
			\addplot  [mark=oplus] table {data/dijkstra-tso-noet-k5.dat};
			\addplot  [mark=triangle] table {data/dijkstra-tso-et-k5.dat};
		\end{axis}
	\end{tikzpicture}

}
\subcaptionbox{dijkstra on PSO ($K_1=5$)\label{Fi:dijkstra-pso-k5}}[.5\linewidth]
{
		\begin{tikzpicture}[scale=.7]
		\begin{axis} [ ymin=0, ymax=600, xmin=2, xmax=40, xlabel=size parameter $n$, ylabel=time(sec), title=dijkstra-pso,
				legend entries={te,fi,robmc,robmc-et},
				legend style= { at={(.35,.95)},anchor=north }
			]
			\addplot  [dashed, mark=x] table {data/dijkstra-pso-te.dat};
			\addplot  [mark=square] table {data/dijkstra-pso-nop.dat};
			\addplot  [mark=oplus] table {data/dijkstra-pso-noet-k5.dat};
			\addplot  [mark=triangle] table {data/dijkstra-pso-et-k5.dat};
		\end{axis}
	\end{tikzpicture}

}
\subcaptionbox{dijkstra on TSO ($K_1=10$) \label{Fi:dijkstra-tso-k10}}[.5\linewidth]
{
	\begin{tikzpicture}[scale=.7]
		\begin{axis} [ ymin=0, ymax=600, xmin=2, xmax=40, xlabel=size parameter $n$, ylabel=time(sec),title=dijkstra-tso,
				legend entries={te,fi,robmc,robmc-et},
				legend style= { at={(.35,.95)},anchor=north }
			]
			\addplot  [dashed, mark=x] table {data/dijkstra-tso-te.dat};
			\addplot  [mark=square] table {data/dijkstra-tso-nop.dat};
			\addplot  [mark=oplus] table {data/dijkstra-tso-noet-k10.dat};
			\addplot  [mark=triangle] table {data/dijkstra-tso-et-k10.dat};
		\end{axis}
	\end{tikzpicture}

}
\subcaptionbox{dijkstra on PSO ($K_1=10$)\label{Fi:dijkstra-pso-k10}}[.5\linewidth]
{
		\begin{tikzpicture}[scale=.7]
		\begin{axis} [ ymin=0, ymax=600, xmin=2, xmax=40, xlabel=size parameter $n$, ylabel=time(sec), title=dijkstra-pso,
				legend entries={te,fi,robmc,robmc-et},
				legend style= { at={(.35,.95)},anchor=north }
			]
			\addplot  [dashed, mark=x] table {data/dijkstra-pso-te.dat};
			\addplot  [mark=square] table {data/dijkstra-pso-nop.dat};
			\addplot  [mark=oplus] table {data/dijkstra-pso-noet-k10.dat};
			\addplot  [mark=triangle] table {data/dijkstra-pso-et-k10.dat};
		\end{axis}
	\end{tikzpicture}

}

\subcaptionbox{ChaseLev on TSO ($K_1=5$) \label{Fi:chaselev-tso-k5}}[.5\linewidth]
{
	\begin{tikzpicture}[scale=.7]
		\begin{axis} [ ymin=0, ymax=600, xmin=2, xmax=40, xlabel=size parameter $n$, ylabel=time(sec),title=chaselev-tso,
				legend entries={te,fi,robmc,robmc-et},
				legend style= { at={(.30,.95)},anchor=north }
			]
			\addplot  [dashed, mark=x] table {data/chaselev-tso-te.dat};
			\addplot  [mark=square] table {data/chaselev-tso-nop.dat};
			\addplot  [mark=oplus] table {data/chaselev-tso-noet-k5.dat};
			\addplot  [mark=triangle] table {data/chaselev-tso-et-k5.dat};
		\end{axis}
	\end{tikzpicture}

}
\subcaptionbox{ChaseLev on PSO ($K_1=5$)\label{Fi:chaselev-pso-k5}}[.5\linewidth]
{
		\begin{tikzpicture}[scale=.7]
		\begin{axis} [ ymin=0, ymax=600, xmin=2, xmax=40, xlabel=size parameter $n$, ylabel=time(sec), title=chaselev-pso,
				legend entries={te,fi,robmc,robmc-et},
				legend style= { at={(.30,.95)},anchor=north }
			]
			\addplot  [dashed, mark=x] table {data/chaselev-pso-te.dat};
			\addplot  [mark=square] table {data/chaselev-pso-nop.dat};
			\addplot  [mark=oplus] table {data/chaselev-pso-noet-k5.dat};
			\addplot  [mark=triangle] table {data/chaselev-pso-et-k5.dat};
		\end{axis}
	\end{tikzpicture}

}

\caption{For all experiments : Timeout=$600$ seconds, $K_2$=all pairs of statement (for soundness) }
\label{Fi:stats}
\end{figure}

\subsection{Experimental Setup}

To enable comparison between the different approaches, we implemented all
four algorithms in the same code base, using \cbmc~\cite{wmmcav13} as the
model checker.  \cbmc explores loops until a given bound.  Our
implementation and the benchmarks used are available online at
\url{http://www.cprover.org/glue} for independent verification of our
results.  The tool takes a C program as an input and assertions in the
program as the specification.

\algref{te} closely approximates the approach used in \dfence~\cite{dfence}. 
\algref{wmmprogrepair} resembles the approach used in
\remmex~\cite{remmex-tso,remmex-pso} and a variant of
\memorax~\cite{memorax-paper,memorax-tool}.  We used \minisat~\cite{minisat}
as the SAT solver in \cbmc.  For all four algorithms incremental SAT solving
is used.  The cardinality constraints used in \algref{brepair} and
\algref{obrepair} are encoded incrementally~\cite{cp14}.  Thus, the program
is encoded only once while the ordering constraints are changed in every
iteration using the assumption interface of the solver.  The experiments
were performed on a machine with 8-core Intel Xeon processors and 48\,GB
RAM.
%
%
The $\mathit{increaseStrategy}(k)$ used for algorithms \algref{brepair} and
\algref{obrepair} doubles the bound $k$.

\subsection{Benchmarks}

Mutual exclusion algorithms such as {\em dekker}, {\em
peterson}~\cite{peterson-mutex}, {\em lamport}~\cite{lamport-mutex}, {\em
dijkstra}~\cite{lamport-mutex} and {\em szymanski}~\cite{szymanski-mutex} as
well as {\em ChaseLev}~\cite{chaselev} and {\em Cilk}~\cite{cilk} work
stealing queues were used as benchmarks.  All benchmarks have been
implemented in C using the \TT{pthread} library.  For mutual exclusion
benchmarks, a shared counter was added and incremented in the critical
section.  An assertion was added to check that none of the increments are
lost.  In addition, all the benchmarks were augmented with a parametric code
fragment shown in \figref{paramsize}, which increases the number of innocent
pairs as $n$ is increased.  The parameter~$n$ was increased from $2$ to $40$
with an increment of $2$.  Thus, each benchmark has $20$ parametric
instances, which makes the total number of problem instances for one memory
model $140$.

\subsection{Results}

We ran our experiments for the TSO and PSO memory models for all the
instances with the timeout of $600$ seconds.  From now on, we will refer to
\algref{te} as \TE, \algref{wmmprogrepair} as \FI, \algref{brepair} as
\robmc and \algref{obrepair} as \robmcet.  In our experiments we found that
all algorithms produce the smallest set of fence placement for every problem
instance.  Thus, we will focus our discussion on the relative performance of
these approaches.

\figref{instances} shows the effect of changing the value of the parameter
$K_1$ in \robmc and \robmcet.  Remember that the bound is increased
gradually from $K_1$ to $K_2$.  Here, $K_2$ is always set to the total
number of statement pairs in the program to guarantee soundness.  \TE and
\FI do not have a parameter~$K_1$, and thus, their corresponding plots are
flat.  \figref{instances} shows that \robmc and \robmcet solve far more
instances than \TE and \FI.  The gap is even wider for the PSO memory model,
which allows more reordering, and thus the number of innocent pairs are
significantly higher compared to TSO on the same program.  As expected,
\robmcet performs better, due to the early termination optimization.  The
value of $K_1$ barely affects the number of solved instances.  The moderate
downward trend for the plots as $K_1$ increases suggests that as $K_1$
increases, \robmc tends to behave more and more like \FI.

\figref{st-pairs} shows the increase in the total number of statement pairs
that can potentially be reordered as the parameter $n$ (\figref{paramsize})
increases for the Peterson algorithm.  As expected, the number of pairs
grows quadratically in $n$.  For PSO, the increase is steeper, as PSO allows
more reordering than TSO.  This explains the better performance of the
\robmc approaches on PSO.

The log-scale scatter plot in \figref{scatterk5} compares the run-time of
\robmcet with $K_1=5$ with \FI over all 280 problem instances. 
 \FI times out significantly more often (data points where both time out are
omitted).  Even on the instances solved by both the approaches, \robmcet
clearly outperforms \FI on all but a few instances.  Those instances where
\FI performs better typically have very few innocent pairs.  Note that the
queries generated by \robmcet are more expensive, as our current
implementation uses cardinality constraints to enforce boundedness.  Thus,
it is possible for \FI to sometimes perform better even though it generates
a larger number of queries to the underlying model checker.

The semi-log-scale plot in \figref{queries} gives the number of queries to
the model checker required by the approaches for the peterson algorithm on
TSO.  \TE and \FI generate exponentially many queries to the model checker
as $n$ increases.  By contrast, the number of queries generated by \robmc
and \robmcet virtually remains unaffected by~$n$.  This is expected as the
search is narrow and focussed owing to the bound $k$.

\figref{peterson-tso-k5} and \figref{peterson-pso-k5} give the relative
performance of all the algorithms when the size and number of innocent pairs
increases with the parameter $n$.  All plots show an exponential trajectory,
indicating that \robmc does not fundamentally reduce the complexity of the
underlying problem.  Even though the number of queries required remains
constant (\figref{queries}), each such query becomes more expensive because
of the cardinality constraints.


However, the growth rate for \robmc and \robmcet is much slower compared to
\TE and \FI.  \figref{peterson-tso-k5} and \figref{peterson-pso-k5}
corroborate the claim that \robmc-based approaches perform much better when
there are a significant number of innocent pairs.  For PSO, the performance
gained by using \robmc is even higher, as PSO allows more reordering. 
Similar trends are observed for dijkstra algorithm in
\figrefs{dijkstra-tso-k5}{dijkstra-pso-k5}.  Plots in
\figrefs{dijkstra-tso-k5}{dijkstra-tso-k10} as well as
\figrefs{dijkstra-pso-k5}{dijkstra-pso-k10} show that the performance of
\robmc-based approaches is not highly sensitive to the value of $K_1$ as it
changes from $5$ to $10$.  This is consistent with the observation made from
\figref{instances}.

The performance comparision for the ChaseLev work stealing queue is given in
\figrefs{chaselev-tso-k5}{chaselev-pso-k5}.  Here it can be seen that the
threshold (in terms of innocent pairs) needed for \robmc to surpass other
approaches is higher.  Even for such a case, \robmc still provides
competitive performance when the number of innocent pairs are low.  \robmc
regains its superiority towards the end as the number of innocent pairs
increases.  Thus, even when every individual query is more expensive (due to
the current implementation that uses cardinality constraints to enforce the
bound), \robmc always provides almost equal or better performance for all
the benchmarks.

\section{Concluding Remarks}\label{Se:conclusion}

ROBMC is a new variant of Bounded Model Checking that has not been explored
before.  Our experimental results indicate substantial speedups when
applying ROBMC for the automated placement of fences on programs with few
culprit pairs and a large number of innocent pairs.  In particular, we
observe that the speedup obtained by using ROBMC increases when targeting a
weaker architecture. Thus, ROBMC adds a new direction in  bounded model checking
which is worth exploring further.

\paragraph{Acknowledgement}

The authors would like to thank Vincent Nimal for helpful discussions on
the related work.

\bibliographystyle{splncs03}
\bibliography{biblio}

\end{document}